\documentclass[floatfix,10pt,superscriptaddress,twocolumn,showpacs,showkeys,aps,pre,final]{revtex4-2}

\usepackage{ifpdf,braket}
\usepackage{amsmath}
\usepackage{amssymb}
\usepackage{graphicx}

\usepackage{lineno}

\usepackage{amsfonts}%
\usepackage{amssymb}%
\usepackage{enumitem}%

\usepackage{textcomp}
\usepackage{ifpdf}

\ifpdf
\usepackage[pdftex,
        colorlinks=true,
        pdftitle={Lyapunov exponent in the Vicsek model},
        pdfauthor={Antonio R. de C. Romaguera},
        baseurl={http://www.df.ufrpe.br},
        pdftoolbar=true,
        pdfmenubar=false,
        pdfpagemode=UseOutlines,
        pdfstartview=FitH,
        bookmarksopen=false
        ]{hyperref}
\fi

\newcommand{\bfF}{{\mathbf{F}}}
\newcommand{\bfG}{{\mathbf{G}}}

\newcommand{\bfv}{{\mathbf{v}}}

\newcommand{\bfx}{{\mathbf{x}}}
\newcommand{\bfU}{{\mathbf{U}}}
\newcommand{\bfu}{{\mathbf{u}}}

\newcommand{\rmi}{{\mathrm i}}

\begin{document}

\title{Lyapunov exponent in the Vicsek model}

\author{L.~H.~Miranda-Filho}
\email[]{lucmiranda@gmail.com}
\affiliation{Departamento de F{\'\i}sica, Universidade Federal Rural de Pernambuco,
Rua Manoel de Medeiros, s/n - Dois Irm\~aos, 52171-900 - Recife, Brazil}

\author{T.~A.~Sobral}

\affiliation{Instituto Federal de Educa\c{c}\~{a}o, Ci\^{e}ncia e Tecnologia do Rio Grande do Norte,
RN 288, s/n, Nova Caic\'{o}, 59300-000 - Caic\'{o}, Brazil}

\author{A.~J.~F.~de Souza}

\affiliation{Departamento de F{\'\i}sica, Universidade Federal Rural de Pernambuco,
Rua Manoel de Medeiros, s/n - Dois Irm\~aos, 52171-900 - Recife, Brazil}

\author{Y.~Elskens}

\affiliation{Aix-Marseille universit\'e, UMR 7345 CNRS, 
Physique des interactions ioniques et mol\'eculaires, 
campus Saint-J\'er\^ome, case 322, av.\ esc.\ Normandie-Niemen, 
FR-13397 Marseille cedex 20, France, EU}

\author{Antonio R.~de C.~Romaguera}
\homepage[]{www.df.ufrpe.br}
\affiliation{Departamento de F{\'\i}sica, Universidade Federal Rural de Pernambuco,
Rua Manoel de Medeiros, s/n - Dois Irm\~aos, 52171-900 - Recife, Brazil}

\date{\today}

\begin{abstract}
The well-known Vicsek model describes the dynamics of a flock of self-propelled particles (SPPs). 
Surprisingly, there is no direct measure of the chaotic behavior of such systems. 
Here, we discuss the dynamical phase transition present in Vicsek systems in light of the largest Lyapunov exponent (LLE),
which is numerically computed by following the dynamical evolution in tangent space for up to two million SPPs. 
As discontinuities in the neighbor weighting factor hinder the computations, we propose a smooth form of the Vicsek model.
We find a chaotic regime for the collective behavior of the SPPs based on the LLE. The dependence of LLE with the applied noise, used as a control parameter, changes sensibly in the vicinity of the well-known transition points of the Vicsek model.
\end{abstract}

\maketitle

\section{Introduction}

Active matter systems are composed of units with a certain ability to absorb energy from their surroundings and, in some way, transform it into mechanical work.
Examples include macroscopic living systems \cite{Eloy2018} as well as microscopic ones \cite{Chen:2017aa}, artificial self-propelled particles~\cite{palacci2015}, and Brownian motors~\cite{astumian2002}, to cite only a few. 
As each unit is autonomous to manage its own energy expenditure, each one evolves according to individual dynamical rules and can respond differently to a given external stimulus.
This makes the description of the behavior of a single entity a very tough task.
Notwithstanding, it is possible to observe the development of intriguing large-scale spatial and temporal patterns during the dynamical evolution of many such units moving independently \cite{Chen:2017aa,Brown:2020aa}.
Such cooperative behavior presents great stability and constitutes the subject of different areas, such as physical chemistry~\cite{sumino2008self}, biology~\cite{deutsch2012collective}, and economics~\cite{anderson2018economy}.

The paradigmatic Vicsek model~\cite{vicsek95} is able to simulate the motion of several forms of active matter, including that of flocks of birds \cite{Brown:2020aa} and schools of fish \cite{Calovi_2014}. 
The model consists of a number of self-driven entities that move together by aligning their velocities with the average velocity of their neighbors.
At densities high enough and noises below a certain value, one observes the phenomenon of clusterization, an ordered motion of the flock, which can unveil different
spatial patterns. Besides the homogeneous polar order,
the model also presents a density bands order~\cite{ginelli2015} and
a so-called cross-sea phase~\cite{prl188003}. The dynamical phase
transitions that occur in the model are first-order~\cite{gregoire2004onset}.
The Vicsek model is probably the most natural starting point to investigate collective dynamics, and so it has been extensively studied~\cite{ginelli2015, chate2008,bolley2012}.
However, we are not aware of any investigation of the emergence of chaos in the Vicsek dynamics. 
Note that a chaotic regime may favor the dispersion of the flock due to sensitivity to small perturbations. 
In addition, either a strongly ordered or a deeply chaotic regime prevent the system from responding quickly to sudden changes in the environmental conditions. 
Thus, it is important to analyze the existence of a chaotic regime in  the Vicsek model. 
The Lyapunov exponent measures how chaotic the dynamics is, as a function of the control parameter \cite{vulpiani2009chaos, miranda21}.

In this article, the largest Lyapunov exponent (LLE) is numerically  computed for the Vicsek model by following the dynamical evolution in tangent space. 
This method requires the integration of the linearized equations of motion, which present discontinuities since 
there is an interaction between two agents only if the distance between them is less than a given cooperation radius.
To get rid of this discontinuity, we introduce a smoothing parameter $\varepsilon$ such that the interaction goes to zero continuously.
The jump discontinuity is recovered for $\varepsilon \to 0$.
Therefore, we systematically study the effect of non-zero values of the smoothing parameter on the continuous form of the Vicsek model.

Section~\ref{sec:model} provides a brief introduction to the Vicsek model.
Sec.~\ref{sec:numdet} establishes the conditions in which the Lyapunov exponents are numerically computed for discrete time dynamics. 
We present suitable linearized equations of motion in Sec.~\ref{sec:linear} to estimate the LLE for the continuous form of the Vicsek model.
The results are presented in Sec.~\ref{sec:results}, 
and finally we dedicate Sec.~\ref{sec:conclusion} to highlighting some conclusions and perspectives.

\section{The Vicsek model}
\label{sec:model}

In his model, proposed in 1995, Vicsek \cite{vicsek95} considers the motion of $N$ self-propelled particles with constant speed $v$ that move in the plane inside a square-shaped cell of size $L$ under periodic boundary conditions. 
Here, we follow the updating scheme of most studies of Vicsek-style models~\cite{ginelli2015, chate2008}, by taking  
\begin{equation}
\bfx_\ell(t+\Delta t) = \bfx_\ell(t) + \bfv_\ell(t+\Delta t) \, \Delta t, 
\label{eq:vicsekposition}
\end{equation}
where one can attribute to the velocity $\bfv_{\ell}$ a propagation angle $\theta_{\ell}$. 
In the Vicsek model, the self-propelled particles line up with their neighbors within an interaction distance, usually taken as a standard distance unit.
The advance angle  $\theta_\ell$ of the particles evolves by consulting those of their neighbors within the cooperation radius, according to the dynamical rule
\begin{equation}
\theta_\ell(t+\Delta t) = \langle \theta_\ell(t) \rangle_r + \Delta \theta_\ell (t).
\label{eq:vicsekangle}
\end{equation}
The term $\langle \theta_\ell(t) \rangle_r$ is related to local alignment mechanisms and denotes the average direction of the velocities of particles in the neighborhood of particle $\ell$, including the particle itself. 
The index $r$ labels all particles inside the interaction range of particle~$\ell$.
The contribution $\Delta \theta_\ell$ plays the role of noise in this model. 
It assumes random values uniformly distributed in the interval $\Delta \theta_\ell \in [-\eta/2, \eta/2]$, where the control parameter of the system $\eta$ is the amplitude of the noise.

The ordering level $\varphi$ constitutes the order parameter in this particular system and is evaluated by
\begin{equation}
\varphi = \frac{1}{Nv}\left| \sum_{\kappa=1}^N{\mathbf{v}_\kappa} \right|,
\label{eq:orderparameter}
\end{equation}
that is, $\varphi$ is the speed of the center of mass divided by the speed $v$ of each particle. 
The ordering level is a normalized quantity, where $\varphi \to 1$ indicates a completely ordered regime, where all particles share the same orientation; while $\varphi \to 0$ indicates a completely non-ordered regime, 
where there is no collective motion. 

From random initial conditions, the global behavior of the self-propelled particles experiences a transient regime in time until assuming a steady order parameter $\varphi^*$ for some values of the noise amplitude $\eta$.

Numerical evidence~\cite{vicsek95,ginelli2015} indicates the existence of a noise amplitude $\eta_c$,
where the system suffers a transition from an ordered regime to a disordered one.  
This reveals the noise amplitude $\eta$ as a control parameter of this dynamical phase transition.
For small values $0 \leq \eta < \eta_c$, the level of ordering $\varphi$ in the steady state is non-zero, 
which reflects the alignment of most agents in a spontaneously chosen direction.
For higher values $\eta_c \leq \eta \leq 2\pi$, the level of ordering $\varphi$ in the steady state goes to zero, owing to the lack of alignment in any direction.
Unlike what was originally claimed~\cite{vicsek95}, 
this is a first-order phase transition~\cite{gregoire2004onset}. 

\section{Numerical determination of the Lyapunov exponent}
\label{sec:numdet}

From the point of view of dynamics, exponential separation of trajectories departing from infinitesimally nearby points is the main characteristic manifested by chaotic regimes \cite{strogatz2001, pikovsky2016}.
In this sense, Lyapunov exponents are measures of the stability of the trajectories \cite{monteiro2002} whose estimations require using both the equations of motion and their local deviations. 
Standard approaches \cite{shimada1979, benettin1980, wolf1985} are based on a linear analysis and consist of observing how infinitesimal perturbations develop around a typical fixed trajectory.

For a generic discrete-time dynamical system, the coordinates $\bfU \equiv (q_1, q_2, \dots, q_\nu)$ evolve according to a recursive relation
\begin{equation}
\bfU(t+\Delta t) = \bfF( \bfU (t) ).
\label{eq:motion}
\end{equation}
The infinitesimal variations $\bfu \equiv (\delta q_1, \delta q_2, \dots, \delta q_\nu)$ follow the linear transformation
\begin{equation}
 \bfu(t+\Delta t) = \frac{\partial \bfF (t)}{\partial \bfU}\, \bfu(t) := \bfG (t) \, \bfu(t),
\label{eq:deviation}
\end{equation}
with $\bfG$ being the Jacobian matrix. 
The perturbation $\bfu (t)$ of trajectory $\bfU (t)$ is 
\begin{eqnarray}
\bfu (t) &=& \bfG (t-\Delta t)\,\bfu(t-\Delta t) \nonumber \\
&=& \bfG (t-\Delta t)\,\bfG (t-2\Delta t)\,\bfu(t-2\Delta t) \nonumber \\
&=& \prod_{n=0}^{s - 1} \bfG (n \Delta t)\,\bfu(0), 
\label{eq:sol_perturbations}
\end{eqnarray}
with $s = t/\Delta t$.
When the elements of $\bfG$ are bounded functions of $t$, the solutions of (\ref{eq:sol_perturbations}) do not grow faster than an exponential function. 
The Lyapunov exponents are defined by
\begin{equation}
\lambda = \lim_{t \to \infty}\frac{1}{t}\ln\frac{\left \| \bfu (t) \right \|}{\left \| \bfu (0) \right \|}.
\label{eq:lyapunov_def}
\end{equation}
Once the limit above exists, the definition (\ref{eq:lyapunov_def}) may be seen as a long-time average of the logarithm of the deviations $\mathbf{u}$ evaluated along a trajectory $\mathbf{U}(t)$. 
Thanks to Birkhoff's and Oseledets' theorem \cite{dorfman1999}, the values of the Lyapunov exponent do not depend on the initial conditions for ergodic processes, except for a set with null measure. 

From a generic initial condition $\bfU_0$, we obtain a reference trajectory $\bfU(t)$. 
For the deviation, a set of independent normalized vectors $(\bfu_1^{(0)}, \bfu_2^{(0)}, \dots, \bfu_\nu^{(0)})$ is defined as initial conditions. 
Both difference Eqs.~(\ref{eq:motion}) and (\ref{eq:deviation}) are iterated, in such a way that in every time step, the values of $\bfU (t)$ generate new elements of the Jacobian matrix, which in turn provides the parameters of the linearized equation. 
After a sufficiently long time, the set of resulting deviation vectors unveil the degree of divergence (or convergence) of the characteristic directions on the state space, each one of them related, according to (\ref{eq:lyapunov_def}), to a value of the Lyapunov exponent. 
Thereby, for a $\nu$-dimensional space of the vectors $\bfu$, there exists a set composed of $\nu$ exponents, which is called the Lyapunov spectrum \cite{pikovsky2016}. 
The largest Lyapunov exponent (LLE) determines the kind of stability of a given trajectory. Positive values are related to exponential instabilities, and therefore chaos, while a negative LLE implies stability. 

The numerical evaluation of~(\ref{eq:lyapunov_def}) requires some careful considerations. 
Exponential solutions of the linearized equations may become problematic in a long-time analysis. 
In addition, to obtain the full Lyapunov spectrum, it is necessary to ensure that iterated deviations refer to independent directions of the state space. 
These issues are circumvented with the use of the Gram--Schmidt procedure \cite{hassani2002}. 
Periodic interventions in the integration of the system prevent, by the normalization, the divergence of the size of the deviation vectors $\bfu$, and the orthogonalization ensure the calculation of the rate of divergence along linearly independent directions.     
This scheme for discrete problems is fully equivalent to the case of continuous-time systems. For more details about the algorithm used, see \cite{miranda19}. 

\section{Linearized equations of the Vicsek model}
\label{sec:linear}

The advance angle of each self-propelled agent $\theta_\ell$ evolves according to Eq.~(\ref{eq:vicsekangle}), which is explicitly written as
\begin{equation}
\theta_\ell(t+\Delta t) = {\rm Arg}{\left[ \bfv_\ell(t) + \sum_{\kappa\neq\ell}^{N} J_{\ell \kappa } \bfv_\kappa(t) \right]} + \Delta \theta_\ell (t),
\label{eq:vicsekangle2}
\end{equation} 
where the function ${\rm Arg}~{\left[ . . . \right]}$ computes the polar angle from the positive horizontal $x$-axis. 
The argument includes all $N$ agents, but the function $J_{\ell \kappa }$ ensures that only agents within the unit cooperation radius are considered, since
\begin{equation}
J_{\ell \kappa} =
  \begin{cases}
    1,       & \quad \text{if } d_{\ell \kappa} \leq 1 , \\
    0,  & \quad \text{otherwise.}
  \end{cases}
\label{eq:defJ}
\end{equation}
In equation~(\ref{eq:defJ}), $d_{\ell \kappa}(t)$ is the Euclidean distance between agents $\ell$ and $\kappa$.
The velocity of the center of mass of the interaction group can be written as $\bfv_{{\mathrm{CM}},\ell} = (m_{x \ell}, m_{y \ell}) \, v/N_\ell$, where $N_\ell = \sum_\kappa J_{\ell \kappa} = 1 + \sum_{\kappa \neq \ell} J_{\ell \kappa}$, a form that enables us to rewrite Eq.~(\ref{eq:vicsekangle2}) as
\begin{equation}
\theta_\ell(t+\Delta t) = {\rm Arg}~\left[m_{x\ell} + \rmi \, m_{y\ell}\right] + \Delta\theta_\ell
\label{eq:vicsekangle3}
\end{equation}
with the usual representation of vectors in ${\mathbb R}^{2}$ as complex numbers. 
The components are
\begin{eqnarray}
m_{x\ell} (t) &=& \cos{(\theta_\ell)} + \sum_{\kappa\neq\ell}^N{J_{\ell \kappa}\cos{(\theta_\kappa)}}\nonumber\\
m_{y\ell} (t) &=& \sin{(\theta_\ell)} + \sum_{\kappa\neq\ell}^N{J_{\ell \kappa}\sin{(\theta_\kappa)}}.
\label{eq:mag}
\end{eqnarray}
The new directions are defined by~(\ref{eq:vicsekangle3}) and are used to update the positions $\bfx_\ell = (x_\ell,y_\ell)$. Hence, by using~(\ref{eq:vicsekposition}), one has
\begin{eqnarray}
x_\ell(t+\Delta t) &= x_\ell(t) + v\cos{(\,\theta_\ell(t + \Delta t)\,)}\, \Delta t, \nonumber\\
y_\ell(t+\Delta t) &= y_\ell(t) + v\sin{(\,\theta_\ell(t + \Delta t)\,)}\, \Delta t.
\label{eq:vicsekposition2}
\end{eqnarray}
The time step corresponds to the updating of the angles and positions of all particles.

Because we have assumed a constant speed $v$ for agents, the state of the system is denoted by a $3N$-dimensional array $\bfU = (x_1, \dots, x_N$, $y_1, \dots, y_N$, $\theta_1, \dots, \theta_N)$. 
This array is defined in the Vicsek model by computing Eqs.~(\ref{eq:vicsekangle3}),~(\ref{eq:mag}), and~(\ref{eq:vicsekposition2}).
In order to obtain the LLE, we also have to compute the respective deviation vector $\bfu = (\delta x_1, \dots, \delta x_N$, $\delta y_1, \dots, \delta y_N$, $\delta \theta_1, \dots, \delta \theta_N)$.

The position equations~(\ref{eq:vicsekposition2}) in linearized form are
 \begin{eqnarray}
\delta x_\ell(t+\Delta t) &=& \delta x_\ell(t) \nonumber \\
& &- v\sin{(\theta_\ell(t + \Delta t))}\, \delta\theta_\ell(t + \Delta t)\, \Delta t,\nonumber \\
\delta y_\ell(t+\Delta t) &=& \delta y_\ell(t) \nonumber \\
& & + v\cos{(\theta_\ell(t + \Delta t))}\, \delta\theta_\ell(t + \Delta t)\, \Delta t,\hspace{0.5cm}
\label{eq:linearposition}
\end{eqnarray}
and the angle equation~(\ref{eq:vicsekangle3}) leads to
 \begin{equation}
\delta \theta_\ell(t+\Delta t) = \frac{m_{x\ell} \, \delta m_{y\ell} -  m_{y\ell}\, \delta m_{x\ell}}{m^2_{x\ell} + m^2_{y\ell}},
\label{eq:linearangle}
\end{equation}
whose differences in the components~(\ref{eq:mag}) are
\begin{eqnarray}
\delta m_{x\ell} (t) &=& -\sin{(\theta_\ell)}\, \delta \theta_\ell \nonumber\\
& &+ \sum_{\kappa\neq\ell}^N\left[\cos{(\theta_\kappa)} \, \delta J_{\ell \kappa} - J_{\ell \kappa }\sin{(\theta_\kappa)}\, \delta \theta_\kappa\right],\nonumber \\
\delta m_{y\ell} (t) &=& \cos{(\theta_\ell)}\, \delta \theta_\ell \nonumber\\
& &+ \sum_{\kappa\neq\ell}^N\left[\sin{(\theta_\kappa)} \, \delta J_{\ell \kappa} + J_{\ell \kappa }\cos{( \theta_\kappa)} \, \delta \theta_\kappa\right],
\label{eq:linearmag}
\end{eqnarray}
where $\delta J_{\ell \kappa}$ denotes the fluctuations generated by the coupling term of the model, which can be rewritten as
\begin{eqnarray}
  \delta J_{\ell \kappa} (t) = \left( \frac{\partial J_{\ell \kappa}}{\partial d_{\ell \kappa}} \right) \frac{1}{d_{\ell \kappa}} & & [\,(x_\ell - x_\kappa)(\delta x_\ell - \delta x_\kappa) \nonumber \\
  & & + (y_\ell - y_\kappa)(\delta y_\ell - \delta y_\kappa)\,].
\label{eq:deltaJ}
\end{eqnarray}
If we realize $J_{\ell \kappa}$ as a Heaviside step function, its derivative $\partial J_{\ell \kappa}/\partial d_{\ell \kappa}$ would behave as a Dirac distribution, that is, it would be nonzero only at the point $d_{\ell \kappa}=1$. 
As a consequence, the deviations from particles located infinitesimally close to the interaction boundary cannot be quantified by Eq.~(\ref{eq:deltaJ}), since $\delta J_{\ell \kappa}$ is not well defined at these points. 

\begin{figure}[h]
 \includegraphics[width=0.9\linewidth]{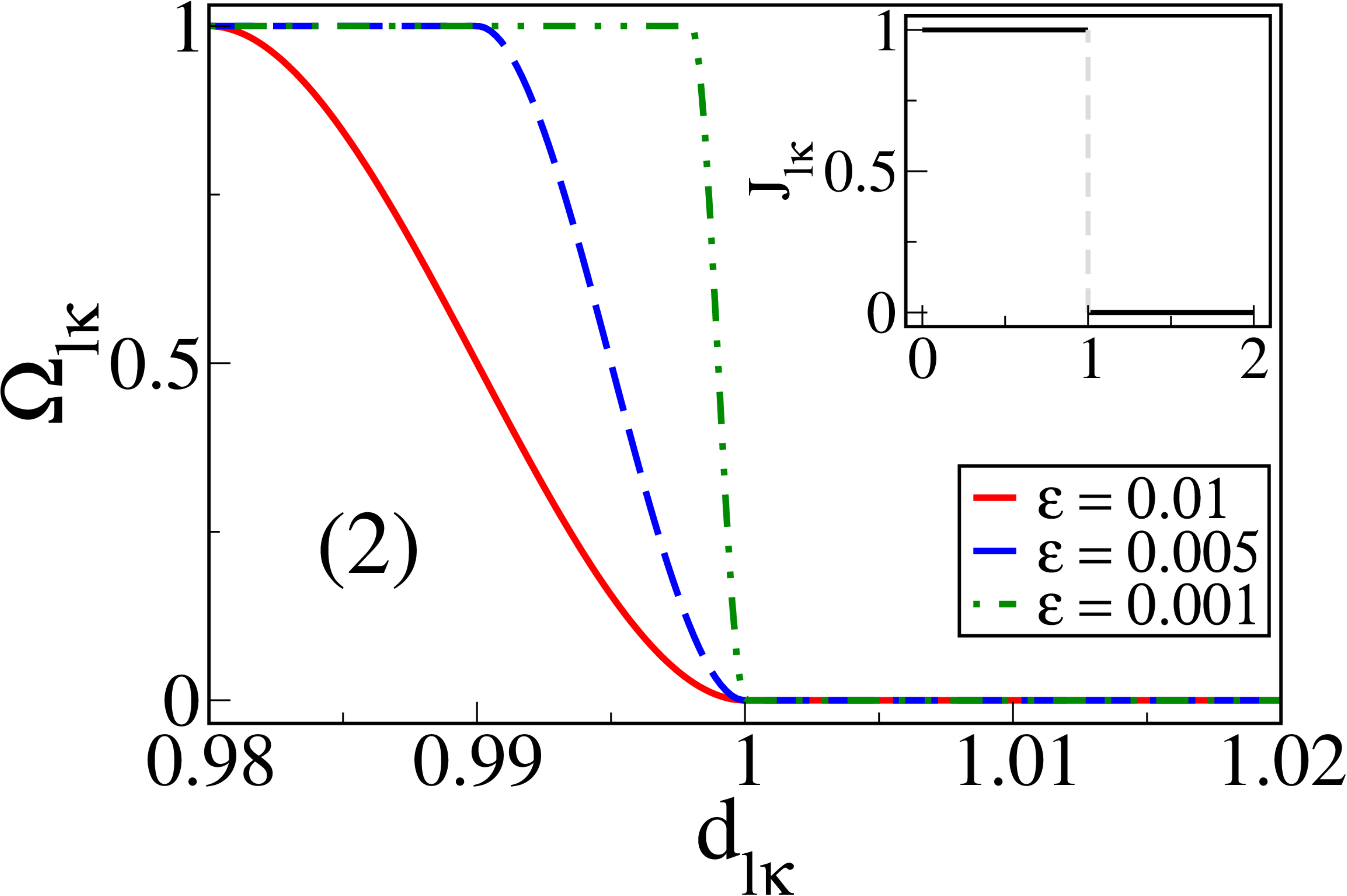}
 \caption{\label{fg:Jgraph}(color online) The weighting factor $\Omega_{\ell \kappa}$ is modeled as a binary function matched with a decreasing cubic function of $d_{\ell \kappa}$. Here $\Omega_{\ell \kappa}$ is plotted for $\varepsilon = \{0.01, 0.005, 0.001\}$. The Vicsek limit corresponds to $\varepsilon \to 0$. Inset: the Vicsek weighting function $J_{\ell \kappa}$ is originally a Heaviside-kind function.}
 \end{figure}

In the present paper, we model $J_{\ell \kappa}$ as a decreasing function $\Omega_{\ell \kappa}$ of $d_{\ell \kappa}$ depending on the smoothing parameter $\varepsilon$. 
Rather than having an abrupt drop, $\Omega_{\ell \kappa}(d_{\ell \kappa},\varepsilon)$ smoothly goes to zero.  
Its mathematical form is
\begin{equation}
\Omega_{\ell \kappa}=\left \{    
\begin{array}{cl}
1 & \text{if $d_{\ell \kappa}<1-2\varepsilon$}\\ \\
{\displaystyle \frac{(d_{lk}-1)^2 (3 \varepsilon+d_{lk}-1)}{4 \varepsilon^3}}& \text{if $1-2\varepsilon \leq d_{\ell \kappa}\leq1$}\\ \\
0 & \text{if $d_{\ell \kappa}>1$}
\end{array}
\right . 
\label{eq:omegadefinition}
\end{equation}
This cubic function is the simplest among all possible choices. 
It has theoretical and numerical advantages over exponential and other similar functions. 
Both the function $\Omega_{\ell \kappa}$ and its  derivatives are continuous, as shown in Fig.~\ref{fg:Jgraph} for three values of $\varepsilon$, and it also recovers the step function in the limit $\varepsilon\to 0$. 
Besides, Eq.~(\ref{eq:omegadefinition}) allows an efficient parallelized CUDA \cite{nvidia2009programming,*scalableCUDA} code making it possible to simulate over $2^{20}$ interacting agents. 
By writing $\Omega_{\ell \kappa}$ in terms of $(d_{lk}-1)$, one keeps the round off errors bounded, and can explore values of the smoothing parameter $\varepsilon$ as small as $10^{-8}$.

The procedure described in Sec.~\ref{sec:numdet} is defined for deterministic dynamical systems. In our case, it works for each given sequence of noise terms $\Delta \theta_\ell (n \Delta t)$ ($1 \leq \ell \leq N$, $0 \leq n \leq s-1$).
As a result, the Lyapunov exponent formally depends on both the initial condition $\bfU_0$ and the realization of the noise. At the limit $t \rightarrow \infty$, the LLE converges to the same value for almost every initial conditions and noise realization. Numerically, one calculates the LLE by averaging over different initial conditions and several noise realizations.  

\section{Results and discussion}
\label{sec:results}

\begin{figure}[!bh]
\includegraphics[width=0.9\linewidth]{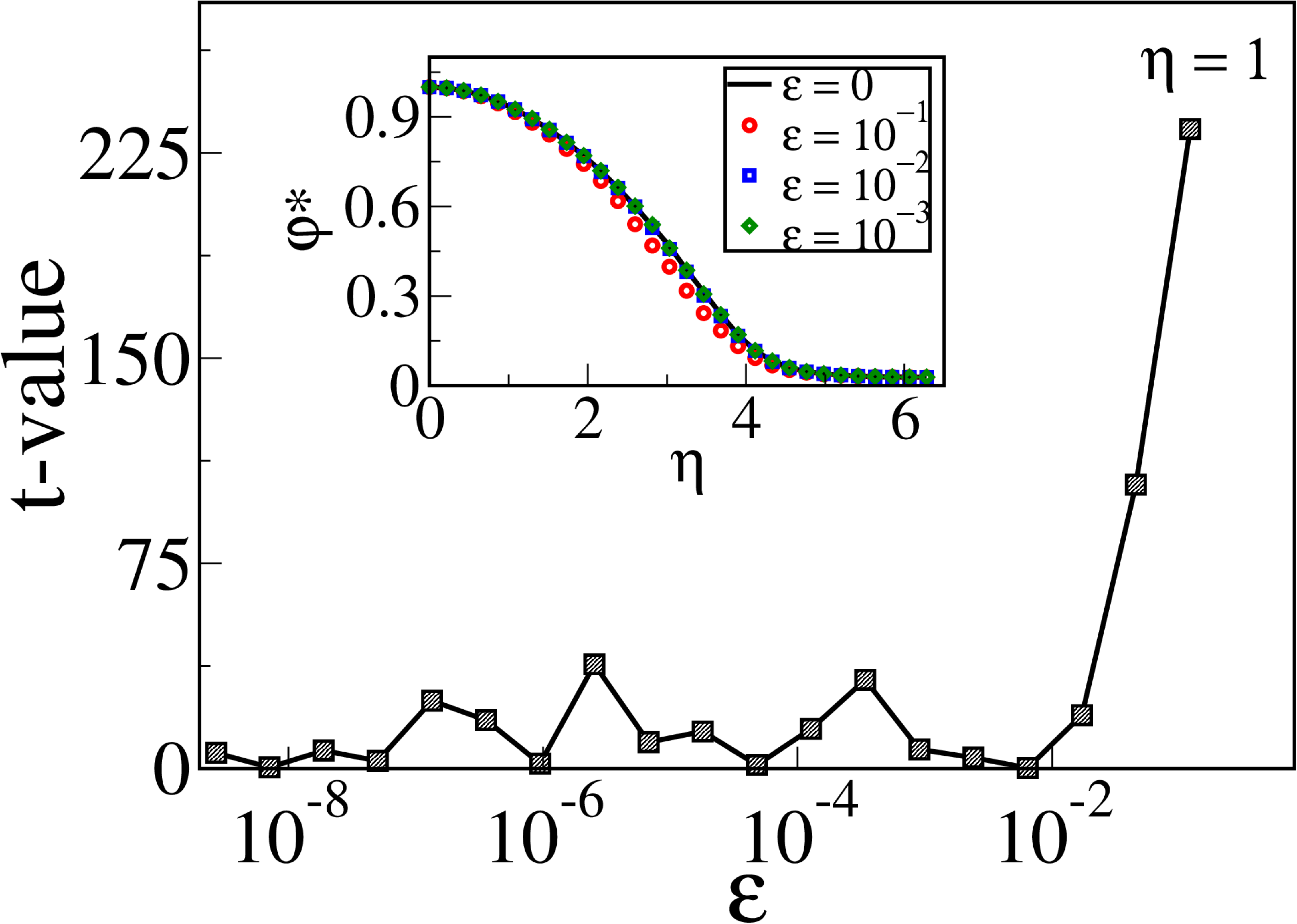}
\caption{\label{fg:ttest} Results of statistical comparisons between time series of the order parameter $\varphi^\ast$ with the Vicsek factor $J_{\ell \kappa}$ and the smoothed weighting factor $\Omega_{\ell \kappa}$. The inset shows the mean level of ordering in the steady state $\varphi^\ast$ as a function of the control parameter $\eta$, for different $\varepsilon$. Data for $L = 16$ and $N = 1\,024$ and $\eta=1.0$.}
\end{figure}

In order to determine the Lyapunov exponents, we iterate both nonlinear and linearized difference equations, defining conveniently the units of space and time with $\Delta t = r = 1$. 
The speed considered, $v = 0.03$, is within the interval suggested by Vicsek~\cite{vicsek95}. 
The noise and the initial conditions are generated by a random number  routine with different seeds.  
The data of the order parameter $\varphi$ extracted from the Vicsek model are compared with those obtained via our particular choice for $\Omega_{\ell \kappa}$.
We use Student's $t$-test (Fig.~\ref{fg:ttest}) for assessing the statistical significance of the difference between the two time series during the stationary state, when the order parameter exhibits an $\eta$-dependent value $\varphi^\ast$. 
For the same set of parameters, a sample of the order parameter $\varphi^\ast$ obtained with the interaction term $J_{\ell \kappa}$ is confronted with samples obtained with the smoothed function $\Omega_{\ell \kappa}$ for different values of $\varepsilon$.
Fig.~\ref{fg:ttest} shows how the two distributions become statistically equivalent as $\varepsilon$ decreases.

The inset of Fig.~\ref{fg:ttest} shows the level of ordering in the steady state $\varphi^\ast$ as a function of
$\eta$, the control parameter of the model. 
For a given $\eta$, the level of ordering tends to be slightly lower for larger $\varepsilon$, which may be directly associated with the slight reduction of the strength of the interaction within the region $1-2\varepsilon < d < 1$, see Eq.~(\ref{eq:omegadefinition}) and Fig.~\ref{fg:Jgraph}.
In addition, the behavior of $\varphi^*$ for the system in the ordered phase $\eta < 1$ and the non-ordered phase $\eta > 5$ are insensitive to the value of $\varepsilon$.
Moreover, as expected, the continuous form of the model reproduces the results of the Vicsek system as $\varepsilon \to 0$.

\begin{figure}[h]
\includegraphics[width=0.9\linewidth]{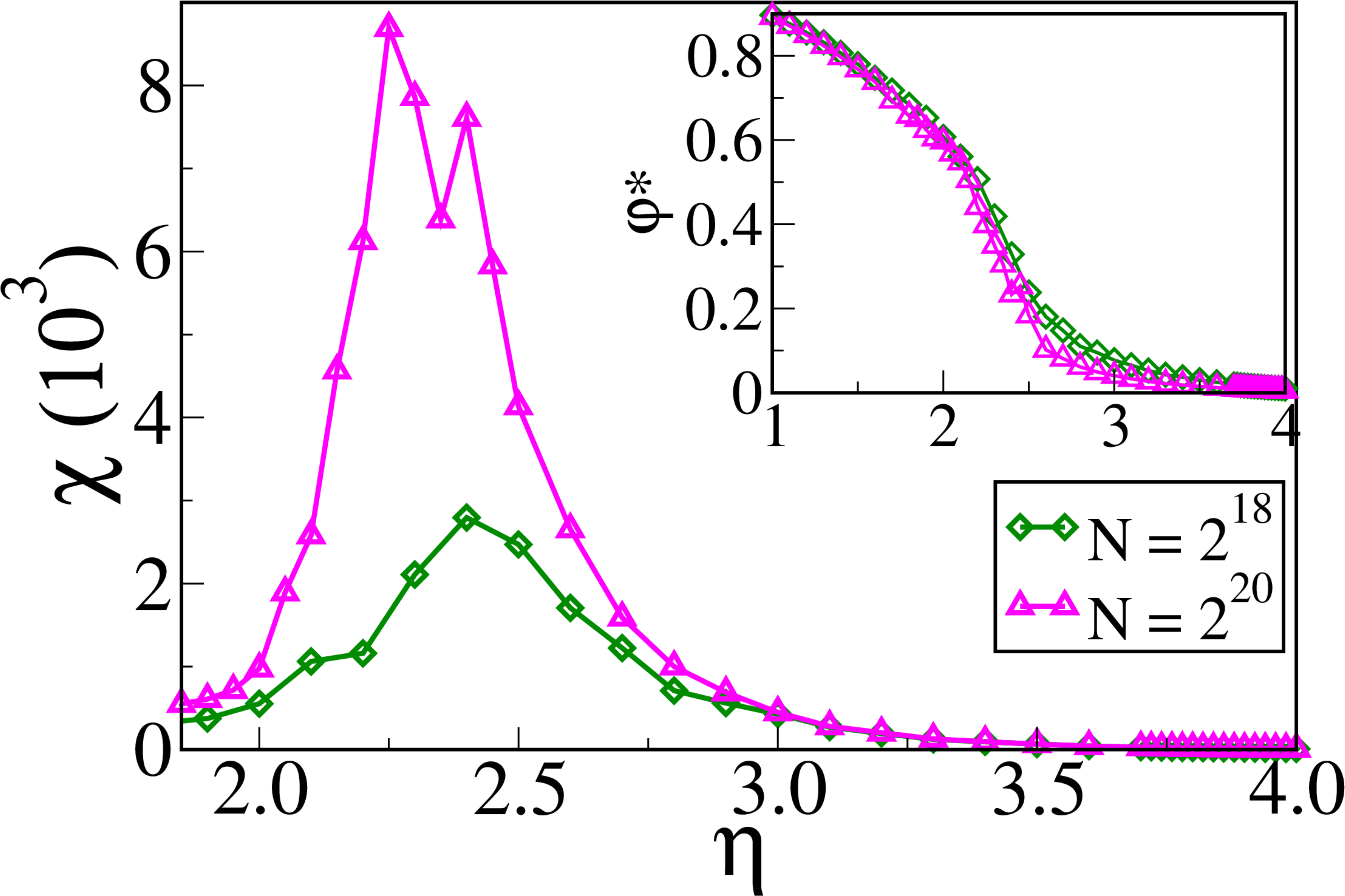}
\caption{ \label{fg:sig} (color online) Susceptibility $\chi$ and order parameter in the steady state $\varphi^\ast$ (inset) as a function of the amplitude noise $\eta$. Data for $L = 256$
and $L = 512$, corresponding to $2^{18}$ and $2^{20}$ particles, respectively.
Here  $\varepsilon = 10^{-8}$.  
}
\end{figure}

The standard Vicsek model presents a polar order for a fixed density, $\rho = N/L^2$,
and $\eta \le \eta_c$ \cite{prl188003}.  
In Fig.~\ref{fg:sig} we show the order parameter and its fluctuation as a function of $\eta$ for $\rho = 4$, $\varepsilon = 10^{-8}$,
and two values of $N$, viz., $N = 2^{18}$ and $N = 2^{20}$.
The fluctuation ($\propto$ susceptibility) is given by $\chi = N \langle \left( \varphi - \langle \varphi \rangle \right)^2 \rangle$, where $\langle \cdots \rangle$ indicates time average in the steady state.
The susceptibility for both sizes exhibit a peak. 
The position of the maximum for the susceptibility is a rough estimate of the phase transition point. From the data for $N=2^{20}$,
we infer that the polar order ceases at $\eta_c \approx 2.25(5)$.
The polar order parameter also drops quickly to an $N$-dependent value for 
$\eta > \eta_c$ as shown in the inset of Fig.~\ref{fg:sig}.

By taking random initial conditions $\bfU_0$ and $\bfu_0$, iterations of Eqs.~(\ref{eq:vicsekposition2}) and~(\ref{eq:vicsekangle2}) provide the reference trajectories, along which the deviation vectors evolve according to the linearized equations of motion~(\ref{eq:linearposition}) and~(\ref{eq:linearangle}). 
After each time step, the Jacobian matrix related with the nonlinear version of the equations is updated, by defining new parameters for the deviations. 
The LLE is calculated from time averages in the stationary state. 
In most of our calculation, $200\,000$ time steps for relaxation time and also for evaluation of the numerical measurements turned out to be enough for convergence of the LLE. As the LLE depends on the particular realization of the noise, we repeated the whole process a certain number of times to calculate the statistical error. For instance, by considering $60$ samples, we estimate an error of about $10^{-3}$ for the system with $N = 1\,024$ particles.
 
\begin{figure}[t]
\includegraphics[width=0.9\linewidth]{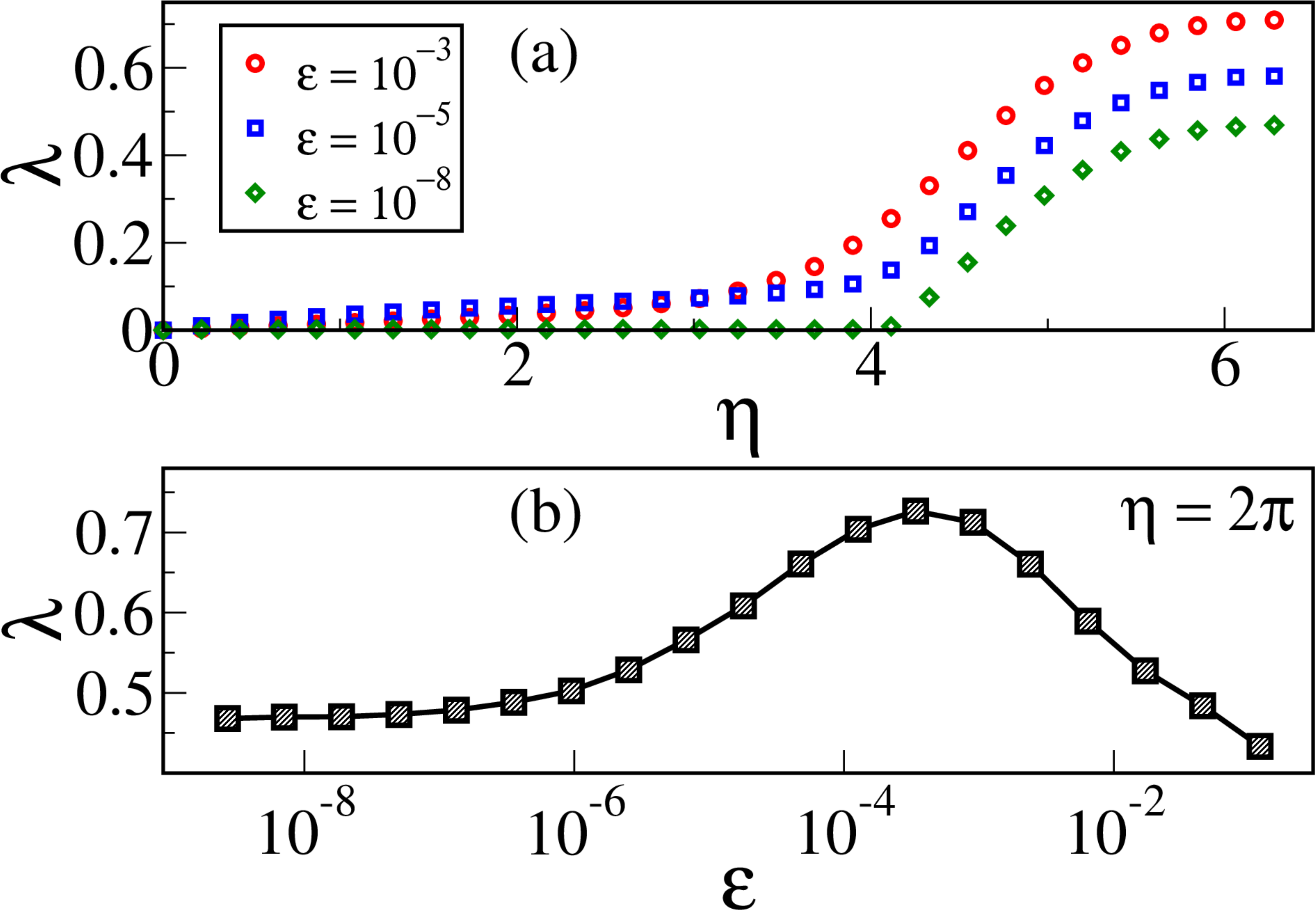}
\caption{\label{fg:lambda} (color online)  (a) LLE as a function of $\eta$ for a few values of $\varepsilon$. (b) The Lyapunov exponent for $\eta=2\pi$ as $\varepsilon$ approaches the Vicsek limit. Results obtained for $L=16$ and $N = 1\,024$.}
\end{figure}

Figure~\ref{fg:lambda}(a) shows how the LLE varies as 
a function of the control parameter $\eta$ for a few chosen values of $\varepsilon$. 
In all cases, the LLE is zero for the deterministic case $\eta=0$, reaching a maximum value when $\eta = 2\pi$.
It is tempting to compare the LLE with the steady state values $\varphi^*$ in the inset of Fig.~\ref{fg:sig}. 
For small amplitudes of the noise, the level of ordering $\varphi^*$ is non-zero in the polar phase, and the LLE indicates little or no exponential disparity between similar trajectories. 
For $\eta$ large, the level of ordering $\varphi^*$ tends to zero in the non-ordered phase, and the LLE indicates exponential divergence.
However, the size of the system is rather small as compared to those in Fig.~\ref{fg:sig} to draw any reliable conclusion.

Fig.~\ref{fg:lambda}(b) shows the maximum value of the LLE for $\eta=2\pi$ as a function of $\varepsilon$. 
As $\varepsilon$ decreases, the LLE increases to a maximum value, after which it decreases and stabilizes at $\lambda_{\varepsilon \to 0}$, a value close to, but less than 0.5. We find that the LLE is a continuous function of $\varepsilon$. Therefore, the value found corresponds to the LLE of the Vicsek model, that is, $\lambda_{\rm Vicsek} = \lambda_{\varepsilon \to 0}$.
From Fig.~\ref{fg:lambda}(b), we expect that $\lambda(2\pi)$ does practically not depend on $\varepsilon$ when $\varepsilon< 10^{-7}$. 
In order to study the thermodynamic limit, we then fixed $\varepsilon = 10^{-8}$ and performed simulations for up to $2^{20}$ agents.

\begin{figure}[h]
\includegraphics[width=0.9\linewidth]{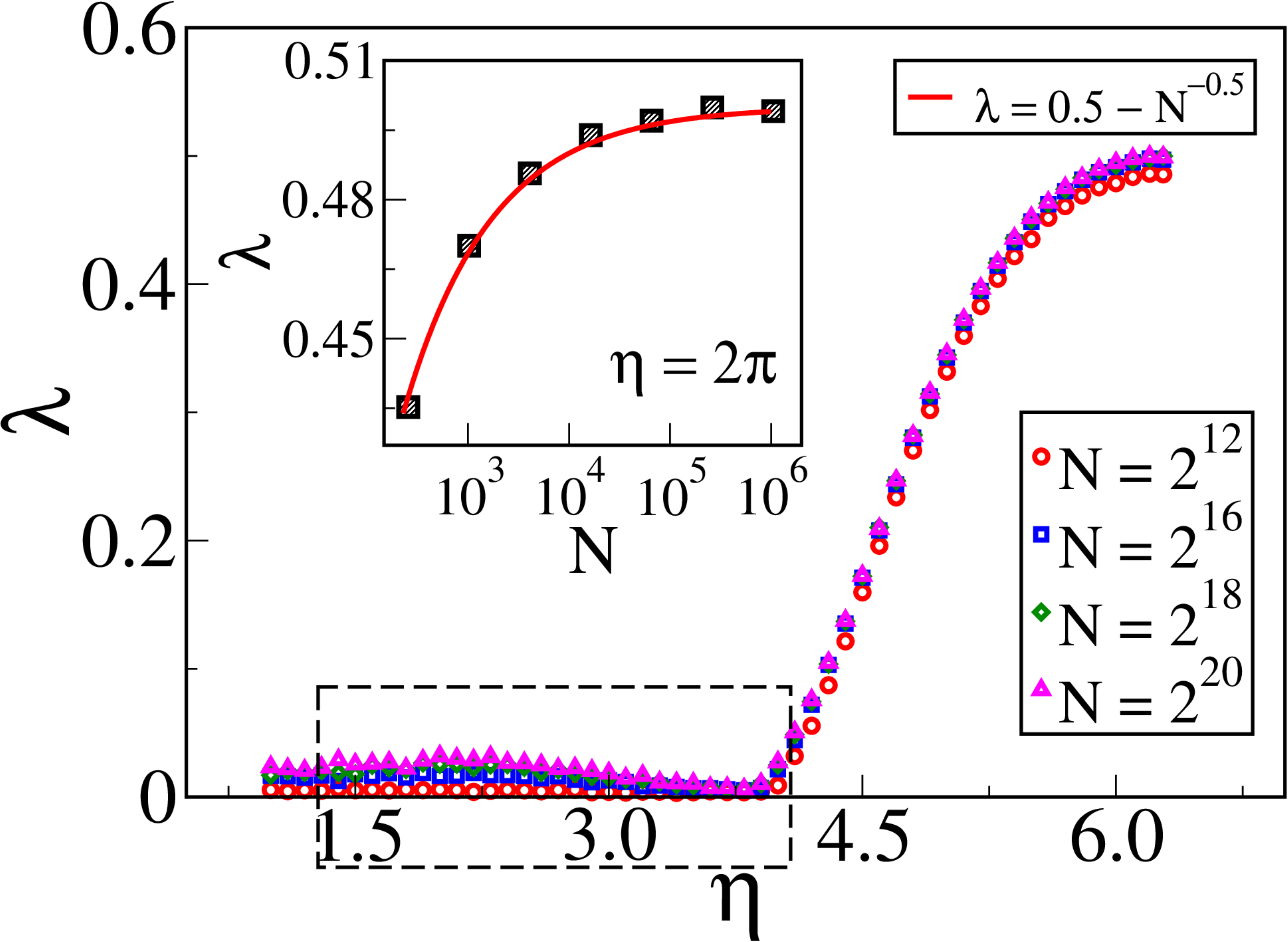}
\caption{\label{fg:N_lambda}(color online) Largest Lyapunov exponent $\lambda$ as function of the noise amplitude $\eta$ for a set of number of particles $N$ with a fixed density $\rho=4$.  
Inset: for $\eta=2\pi$, the largest Lyapunov exponent increases monotonically with the number of particles. 
Dotted region shows finite size dependence.
All data are obtained for $\varepsilon = 10^{-8}$.}
\end{figure}

The impact of the number of particles $N$ on the LLE $\lambda$ is shown in Fig.~\ref{fg:N_lambda} for a fixed density $\rho = 4$.
The general form for $\lambda(\eta)$ is similar to those found in Fig.~\ref{fg:lambda}(a) for fewer particles.
In the inset of Fig.~\ref{fg:N_lambda}, we show $\lambda$ as a function
of the number of particles for $\eta=2\pi$, where $\lambda$ reaches its
maximum. 
One sees that $\lambda$ increases monotonically with $N$ and
seems to saturate very closely to $1/2$. 
The data points behave very well as $(1/2 - \lambda) = N^{-1/2}$, as illustrated by the continuous line in the inset of Fig.~\ref{fg:N_lambda}, which allows extrapolating to the thermodynamic limit $\lambda_{\infty} = 1/2$.
It is also noticeable in Fig.~\ref{fg:N_lambda} that the LLE presents some dependence on $N$ for $\eta < 4$. It is apparent from the figure that the value of noise separating the chaotic phase from that of small chaos takes place at $\eta$ slightly below $4.0$, which is well above
$\eta_c = 2.25$. Also, although hardly visible in the scale of Fig.~\ref{fg:N_lambda}, it is possible
to notice a bump near $\eta = \eta_c$, see highlighted region in it. So we decide to have a closer look at $\eta$ in
the range $1.5 \le \eta \le 4.0$.

\begin{figure}[h]
\includegraphics[width=0.9\linewidth]{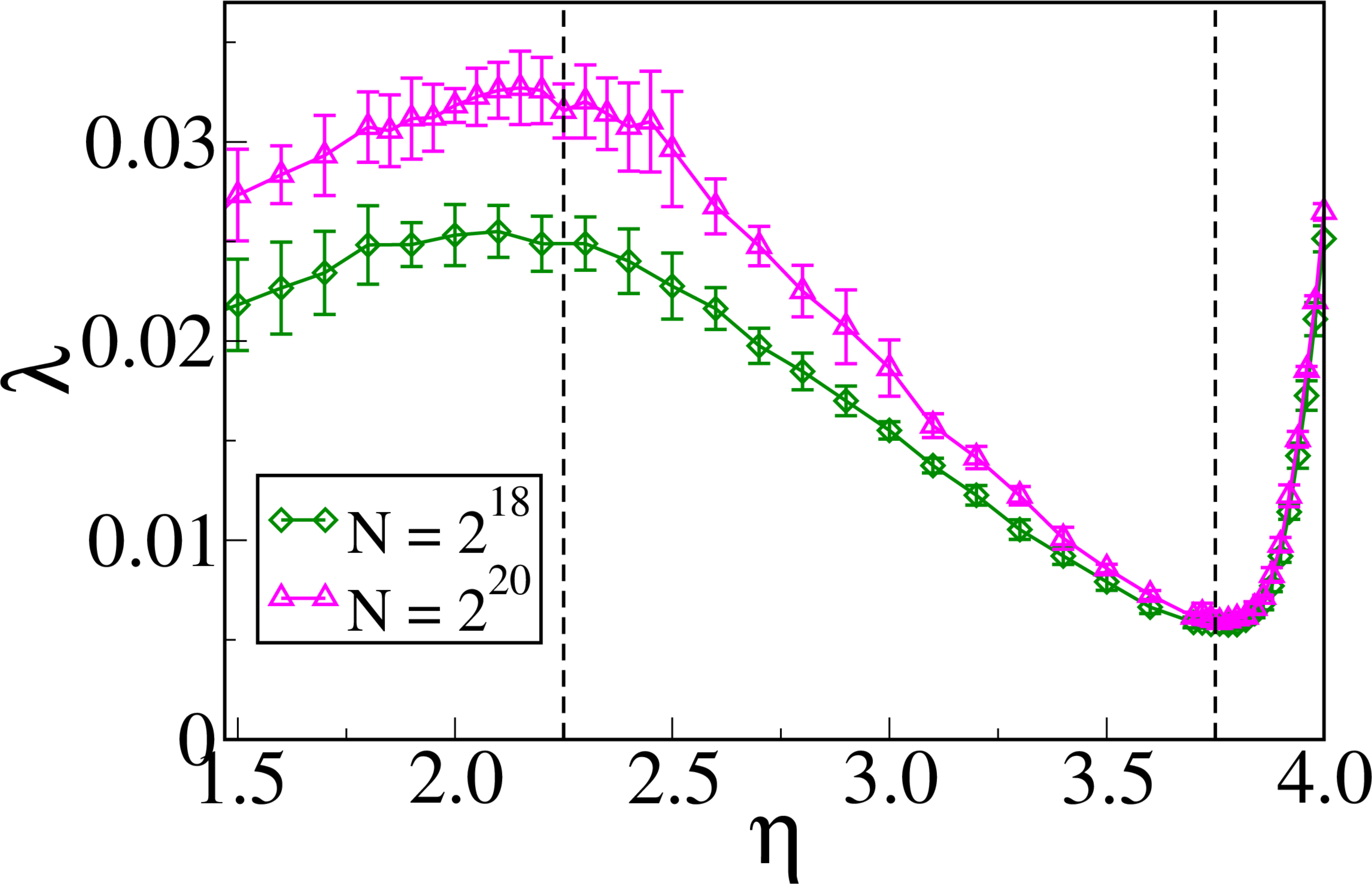}
\caption{
\label{fg:new}(color online) LLE $\lambda$ as function of $\eta$ for $N=2^{18}$, 
$N=2^{20}$, with a fixed density $\rho=4$ and $\varepsilon = 10^{-8}$. 
Vertical lines mark the transition points $\eta_c = 2.25$ and $\eta_d = 3.78$.
The symbols correspond to an average over $30$ independent runs of $2\times 10^5$ time steps.    
}
\end{figure}

Figure \ref{fg:new} presents in more detail
the dependence of $\lambda$ as a function of $\eta$
in the region where it showed finite-size effects
and a noticeable change of behavior.
We show results for $N=2^{18}$ and $N=2^{20}$
with each data-point corresponding to an average
over $30$ independent runs of $2\times 10^5$ time steps.
From the curve of Fig.~\ref{fg:new}, we see that $\lambda$ exhibits a maximum
at $\eta$ close to $\eta_c = 2.25$
where the polar order vanishes.
The LLE decreases almost linearly with $\eta$ for $\eta > \eta_c$ reaching a minimum
at $\eta = 3.78(2)$, from then on a vertiginous growth in chaos occurs.

Interestingly, the changes in the derivative of $\lambda(\eta)$ with respect to $\eta$ coincides with transition points. Therefore, Fig.~\ref{fg:new} displays a novel procedure to find the phase boundaries for the Vicsek model.

Accordingly, Fig.~\ref{fg:new} shows that the standard Vicsek model
experiences two phase transitions for density $\rho = 4$.
The first transition occurs at $\eta_c = 2.25(5)$
where the polar order disappears. The second one
occurs at $\eta_d = 3.78(2)$
marking the beginning of a gas-like phase.
We did not identify the nature of the
intermediate small-chaos phase.
However, one recently discovered
four different phases at very low
density for the Vicsek model \cite{prl188003}.
Based on this finding, we believe that
the intermediate phase corresponds to an 
ordered band phase \cite{ginelli2015}.

The change in the diffusion regime marks the transition between the ordered and disordered phases \cite{gregoire2001active}. 
Thus, the dynamic behavior provides complementary information about the transition point.  
Above $\eta_d = 3.78$, the flock disperses diffusively. 
Below $\eta_d$, the flock has a very distinct behavior. 
For $\eta \ll \eta_d$, the particles move ballistically, owing to the highly ordered collective behavior.
As $\eta$ approaches $\eta_d$ from below, the noise is not strong enough to break the order, but the flock's velocity frequently changes direction. 
Also, the flock breaks apart sometimes. 
The mean square displacement presents a superdiffusive-like motion for
the intermediate time scale before saturating due to 
finite-size effects. In the disordered phase, a diffusive-like
motion emerges right at the transition point and remains
diffusive up to $\eta = 2\pi$, the maximum allowed noise.

When $\eta = 2\pi$, the term $\Delta \theta_\ell$ is a random variable, uniform on $[-\pi, \pi]$, 
drawn independently for each particle. 
Thus, the Vicsek and Random Walk models present equivalent dynamics with, 
however, distinct sensitivity to initial conditions, 
as their deviations obey different linearized equations.
In a Random Walk, the angles are updated according to
\begin{equation}
\theta_\ell(t+\Delta t) = f(\theta_\ell(t)) + \Delta \theta_\ell(t), 
\label{eq:rw}
\end{equation}
where $f$ is the identity function $f(\theta) = \theta$ and
the noise terms $\Delta \theta_\ell(t)$ are the same as in Eq.~(\ref{eq:vicsekangle2}). 
For $\eta = 2\pi$, we have an isotropic Random Walk.
As Eq.~(\ref{eq:linearangle}) is replaced with $\delta \theta_\ell (t + \Delta t) = \delta \theta_\ell (t)$, 
there is no chaos, implying $\lambda = 0$. 
 
On the other hand, for the Vicsek model, 
the relation between the new and old angles are nonlinear by construction, 
see the Eq.~(\ref{eq:vicsekangle2}).
Such nonlinearity also reflects in the form of deviations, as shown by Eqs.~(\ref{eq:linearposition}) and (\ref{eq:linearmag}).  
One computes the Lyapunov exponent by taking the reference trajectory for a given noise and the nearby trajectories with the same noise. As a result, the angles $\theta_\ell$ in (\ref{eq:linearposition}) and (\ref{eq:linearmag}) are independent random, uniform on the circle ($\eta = 2\pi$), but the first variations $\delta \theta_\ell$ are not independent, which culminates in the results of the Figs.~\ref{fg:N_lambda} and \ref{fg:new}.

One relevant issue is whether the LLE can detect phase transitions in other scenarios.
We mitigate this concern by trying different values for the model parameters.
Thus, we run simulations for a few values of the density $\rho$, in which the
speed of the particles is $v = 0.5$ to allow a direct comparison with the
results of Ref. \cite{PhysRevE.92.062111} where is reported a thorough analysis
of the Vicsek model.

\begin{figure}[b]
\includegraphics[width=0.9\linewidth]{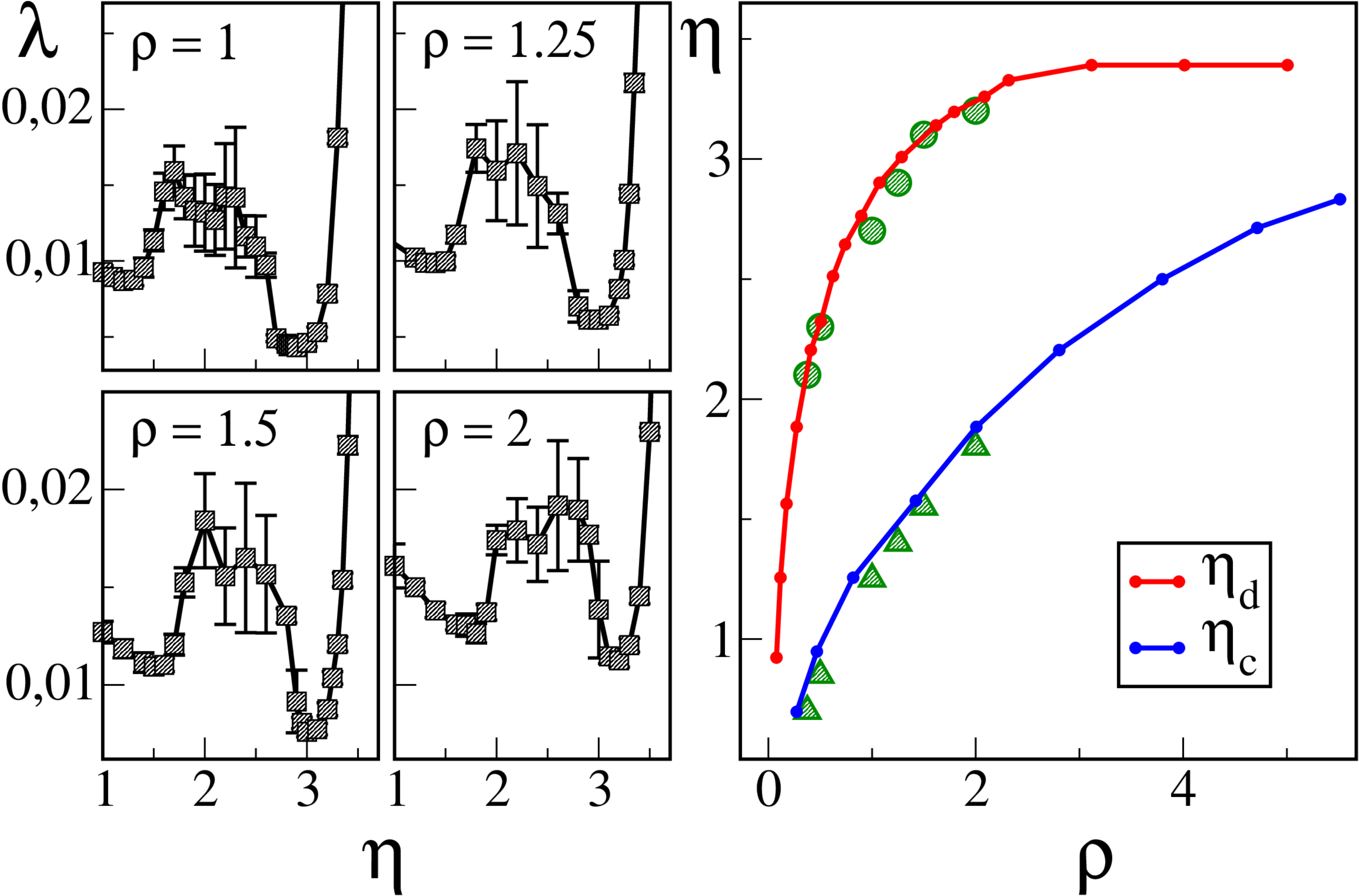}
\caption{
\label{fg:tst}(color online) (Left) LLE $\lambda$ as function of $\eta$ for densities $\rho = 1, 1.25, 1.5, \mbox{and } 2$. (Right) Vicsek model phase diagram for speed $v=0.5$. Solid lines are from Ref.~\cite{PhysRevE.92.062111}, whereas square and triangle symbols correspond to our numerical results for $\eta_c$ and $\eta_d$, respectively. 
}
\end{figure}	

In the left panel of Fig.~\ref{fg:tst}, we show $\lambda$ as a function of $\eta$
for $\rho$ = 1, 1.25, 1.5, 2 with $N$ = 1\,048\,576, 1\,310\,720, 1\,572\,864, and
2\,097\,152, respectively. In all cases shown, one can see two minima separated by a plateau.
That is, $\lambda$ sensibly changes its behavior at two different noise values.
The transition points are roughly located between one of the extremities of the plateau and the nearest minimum. The data presented in the left panel of Fig.~\ref{fg:tst}
show that the LLE can also identify the well-known transition points of the Vicsek model at different densities.

In the right panel of Fig.~\ref{fg:tst}, we report a quantitative comparison between our estimates (square and triangle symbols) for the transition points and the precise data from Ref.~\cite{PhysRevE.92.062111} (solid lines). 
For compatibility with our scale for the noise, we multiply the noise values of the Ref.~\cite{PhysRevE.92.062111} by $2\pi$.  
The transition lines $\eta_c$ and $\eta_d$ are both monotonically increasing functions of the density $\rho$. Below $\eta_c$ (blue) continuous line in the right panel of Fig.~\ref{fg:tst}, the polar order is stable, whereas above the $\eta_d$ (red) line is the gaseous phase.
For $\eta_c(\rho) < \eta <\eta_d(\rho)$, there exists a mixed-phase.   
As can be seen, our results satisfactorily agree with the data already present in the literature.

\section{Concluding remarks}
\label{sec:conclusion}

In the present study, we focused on characterizing the chaotic
behavior of self-propelled agents ruled by the Vicsek model. 
In this context, discontinuities in the dynamical rules hinder
the determination of the Lyapunov exponents. 
To circumvent this obstacle, we introduced a continuous
decreasing function $\Omega(\varepsilon)$ which recovers the
Vicsek dynamics in the limit $\varepsilon \to 0$.

For a sufficiently large number of particles,
the largest Lyapunov exponent (LLE) 
shows peculiar behavior at phase boundaries.
We found that the Vicsek model suffers two phase
transitions as the noise is varied at fixed density $\rho = 4$.
Roughly speaking, the LLE presents a slightly chaotic
regime in the ordered phases and a
substantially chaotic regime as it enters in the gas-like phase.     

Close to the transition points, 
the LLE presents a striking change of behavior. 
The derivative of LLE with respect to noise 
changes its sign at the phase boundary points. 
In particular, the almost linear behavior of 
the LLE turns into a rapid growth right after 
the system enters the disordered phase.

Similar behavior of the LLE has been reported for a
Hamiltonian system  near a second-order phase
transition~\cite{Caiani1997}. We propose that the Lyapunov
exponent can work as a phase transition indicator, also
for this out-of-equilibrium system.

The LLE regulates over long times the exponential growth
rate of the separation between similar trajectories.
Through the LLE, the predictability of Vicsek's dynamical
system has been studied and we found different behaviors
for ordered and non-ordered phases. The full Lyapunov
spectrum and Lyapunov vector analysis are natural next
topics for investigation. Furthermore, it is worth
calculating the LLE in lower densities to investigate
whether it can detect the cross-sea phase recently observed \cite{prl188003}.

\section*{Acknowledgments}

A.~R.~de C.~Romaguera thanks the Brazilian agency FACEPE for financial help (grant No.~APQ-0198-1.05/14). L.~H.~Miranda-Filho thanks the PIIM, CNRS-AMU for their hospitality, Bruno V.~Ribeiro for discussions, and CAPES for financial support. 
We also thank anonymous reviewers for their constructive comments. 
We gratefully acknowledge the support of the Nvidia Corporation with the donation of GPUs for our research.


\hyphenation{physico-chemical}

\end{document}